\def\mn{{Mon.\@ Not.\@ Roy.\@ Ast.\@ Soc.\ }}
\def \jetpl {JETP Lett.\ }
\documentclass[prd,aps,floats,preprintnumbers,preprint]{revtex4}

\usepackage{graphicx}
 
\newcommand{\be}{\begin{equation}}
\newcommand{\ee}{\end{equation}}
\newcommand{\bea}{\begin{eqnarray}}
\newcommand{\eea}{\end{eqnarray}}

\begin{document}

\preprint{YITP-09-76}
\preprint{RESCUE-28-09}

\setlength{\unitlength}{1mm}

\title{Testing homogeneity with galaxy number counts  : light-cone metric and general low-redshift expansion for a central observer in a matter dominated isotropic universe without cosmological constant.}

\author{Antonio Enea Romano}
\affiliation{Yukawa Institute for Theoretical Physics, Kyoto University,
Kyoto 606-8502, Japan}

\affiliation{
Research Center for the Early Universe (RESCEU),
Graduate School of Science, The University of Tokyo, Tokyo 113-0033, Japan}

\begin{abstract}
As an alternative to dark energy it has been suggested that we may be at the center
 of an inhomogeneous isotropic universe described by
 a Lemaitre-Tolman-Bondi (LTB) solution of Einstein's field equations.
In order to test this hypothesis we calculate the general analytical formula to fifth order for the redshift spherical shell mass.
Using the same analytical method we write the metric in the light-cone by introducing a gauge invariant quantity $G(z)$ which together with the luminosity distance $D_L(z)$ completely determine the light-cone geometry of a LTB model.

\end{abstract}

\maketitle
\section{Introduction}
High redshift luminosity distance measurements \cite{Perlmutter:1999np,Riess:1998cb,Tonry:2003zg,Knop:2003iy,Barris:2003dq,Riess:2004nr} and
the WMAP measurement \cite{WMAP2003,Spergel:2006hy} of cosmic
microwave background (CMB) interpreted in the context of standard 
FLRW cosmological models have strongly disfavored a matter dominated universe,
 and strongly supported a dominant dark energy component, giving rise
to a positive cosmological acceleration, which we will denote by $a^{FLRW}$
 (not to be confused with the scale factor $a$).
As an alternative to dark energy, it has been 
proposed \cite{Celerier:1999hp,Nambu:2005zn,Kai:2006ws}
 that we may be at the center of an inhomogeneous isotropic universe 
described by a Lemaitre-Tolman-Bondi (LTB) solution of Einstein's field 
equations, where spatial averaging over one expanding and one contracting 
region is producing a positive averaged acceleration $a_D$.
Another more general approach to map luminosity distance as a function of
 redshift $D_L(z)$ to LTB models has been recently
 proposed \cite{Chung:2006xh,Yoo:2008su},
 showing that an inversion method can be applied successfully to 
reproduce the observed $D_L(z)$.   

The main point is that the luminosity distance is in general sensitive 
to the geometry of the space through which photons are propagating along
null geodesics, and therefore arranging appropriately the geometry of 
a given cosmological model it is possible to reproduce a given $D_L(z)$.
For FLRW models this corresponds to the determination of $\Omega_{\Lambda}$
 and $\Omega_m$ and for LTB models it allows to determine the 
functions $E(r),M(r),t_b(r)$.

Another observable which could be used to constraint LTB models is the redshift spherical shell mass $mn(z)$, which we calculate
for a central observer up to the filth order in the red-shift.
Using the same set of geodesic equations derived to obtain such central expansion we also write the light-cone metric in red-shift space. 

\section{Lemaitre-Tolman-Bondi (LTB) Solution\label{ltb}}
Lemaitre-Tolman-Bondi  solution can be
 written as \cite{Lemaitre:1933qe,Tolman:1934za,Bondi:1947av}
\begin{eqnarray}
\label{eq1} %
ds^2 = -dt^2  + \frac{\left(R,_{r}\right)^2 dr^2}{1 + 2\,E}+R^2
d\Omega^2 \, ,
\end{eqnarray}
where $R$ is a function of the time coordinate $t$ and the radial
coordinate $r$, $R=R(t,r)$, $E$ is an arbitrary function of $r$, $E=E(r)$
and $R,_{r}=\partial R/\partial r$.

Einstein's equations give
\begin{eqnarray}
\label{eq2} \left({\frac{\dot{R}}{R}}\right)^2&=&\frac{2
E(r)}{R^2}+\frac{2M(r)}{R^3} \, , \\
\label{eq3} \rho(t,r)&=&\frac{M,_{r}}{R^2 R,_{r}} \, ,
\end{eqnarray}
with $M=M(r)$ being an arbitrary function of $r$ and the dot denoting
the partial derivative with respect to $t$, $\dot{R}=\partial R(t,r)/\partial t$.
 The solution of Eq.\ (\ref{eq2}) can be expressed parametrically 
in terms of a time variable $\tau=\int^t dt'/R(t',r) \,$ as
\begin{eqnarray}
\label{eq4} Y(\tau ,r) &=& \frac{M(r)}{- 2 E(r)}
     \left[ 1 - \cos \left(\sqrt{-2 E(r)} \tau \right) \right] \, ,\\
\label{eq5} t(\tau ,r) &=& \frac{M(r)}{- 2 E(r)}
     \left[ \tau -\frac{1}{\sqrt{-2 E(r)} } \sin \left(\sqrt{-2 E(r)}
     \tau \right) \right] + t_{b}(r) \, ,
\end{eqnarray}
where  $Y$ has been introduced to make clear the distinction
 between the two functions $R(t,r)$ and $Y(\tau,r)$
 which are trivially related by 
\begin{equation}
R(t(\tau,r))=Y(\tau,r) \, ,
\label{Rtilde}
\end{equation}
and $t_{b}(r)$ is another arbitrary function of $r$, called the bang function,
which corresponds to the fact that big-bang/crunches can happen at different
times. This inhomogeneity of the location of the singularities is one of
the origins of the possible causal separation \cite{Romano:2006yc} between 
the central observer and the spatially averaged region for models
 with positive $a_D$.

We introduce the variables
\begin{equation}
 A(t,r)=\frac{R(t,r)}{r},\quad k(r)=-\frac{2E(r)}{r^2},\quad
  \rho_0(r)=\frac{6M(r)}{r^3} \, ,
\end{equation}
so that  Eq.\ (\ref{eq1}) and the Einstein equations
(\ref{eq2}) and (\ref{eq3}) are written in a form 
similar to those for FLRW models,
\begin{equation}
\label{eq6} ds^2 =
-dt^2+A^2\left[\left(1+\frac{A,_{r}r}{A}\right)^2
    \frac{dr^2}{1-k(r)r^2}+r^2d\Omega_2^2\right] \, ,
\end{equation}
\begin{eqnarray}
\label{eq7} %
\left(\frac{\dot{A}}{A}\right)^2 &=&
-\frac{k(r)}{A^2}+\frac{\rho_0(r)}{3A^3} \, ,\\
\label{eq:LTB rho 2} %
\rho(t,r) &=& \frac{(\rho_0 r^3)_{, r}}{6 A^2 r^2 (Ar)_{, r}} \, .
\end{eqnarray}
The solution of Eqs.\ (\ref{eq4}) and (\ref{eq5}) can now be written as
\begin{eqnarray}
\label{LTB soln2 R} a(\eta,r) &=& \frac{\rho_0(r)}{6k(r)}
     \left[ 1 - \cos \left( \sqrt{k(r)} \, \eta \right) \right] \, ,\\
\label{LTB soln2 t} t(\eta,r) &=& \frac{\rho_0(r)}{6k(r)}
     \left[ \eta -\frac{1}{\sqrt{k(r)}} \sin
     \left(\sqrt{k(r)} \, \eta \right) \right] + t_{b}(r) \, ,
\end{eqnarray}
where $\eta \equiv \tau\, r = \int^t dt'/A(t',r) \,$ and $A(t(\eta,r),r)=a(\eta,r)$.

In the rest of paper we will use this last set of equations .
Furthermore, without loss of generality, we may set 
the function $\rho_0(r)$ to be a constant,
 $\rho_0(r)=\rho_0=\mbox{constant}$, corresponding to the choice of coordinates in which $M(r)\propto r^3$, and we will call this, following \cite{Nambu:2005zn}, the FLRW gauge.

We need three functions to define a LTB solution, but because of the invariance under general coordinate transformations, only two of them are really independent. This implies that two observables are in principle sufficient to solve the inversion problem of mapping observations to a specific LTB model, for example the luminosity distance $D_L(z)$ and the redshift spherical shell mass $m(z)n(z)=mn(z)$.
As observed by \cite{Celerier:2009sv}, there as been sometime some confusion about the general type of LTB models which could be used to explain cosmological observations, so it is important to stress that without restricting the attention on models with homogeneous big bang , $t_b(r)=0$, a void is not necessary to explain both $D_L(z)$ and $m(z)n(z)=mn(z)$ without cosmological constant.
We call $mn(z)$ redshift spherical shell mass, since this quantity it is not the galaxy number counts as it is called in the interesting paper  \cite{Celerier:2009sv}, but the product of the source number density $n(z)$ times the source mass function $m(z)$, and it has dimension of energy.

This should not be confused with the redshift spherical shell energy $E_{RSS}$ introduced in \cite{Romano:2007zz}, which is a quantity obtained by integrating $mn(z)$ over varying redshift intervals $\Delta Z(z)$ 

\bea
E_{RSS}(z)=\int \limits_{z}^{z+\Delta Z(z)} {4\pi mn(z')d\,z'}\\
t(z)-t(z+\Delta Z(z))=\Delta t
\eea

corresponding to the same constant time interval $\Delta t$, which, if chosen to be sufficiently smaller than the time scale of astrophysical evolution of the source, should eliminate the effect of the source evolution on $mn(z)$.

\section {Geodesic equations}

We will adopt the same method developed in \cite{Romano:2009xw} to find the null geodesic equation in the coordinates $(\eta,t)$, but here instead of integrating numerically the differential equations we will find a local expansion of the solution around $z=0$ corresponding to the point $(t_0,0)\equiv(\eta_0,t)$, where $t_0=t(\eta_0,r)$.
We will also provide more details about the geodesic equation derivation which were presented in \cite{Romano:2009xw} in a rather concise way. 
We will indeed slightly change notation to emphasize the fully analytical r.h.s. of the equations obtained in terms of $(\eta,t)$, on the contrary of previous versions of the light geodesic equations which require some numerical calculation of $R(t,r)$ from the Einstein's equation(\ref{eq2}). 

For this reason this formulation is particularly suitable for the derivation of analytical results.

The luminosity distance for a central observer in a LTB space 
as a function of the redshift is expressed as
\be
D_L(z)=(1+z)^2 R\left(t(z),r(z)\right)
=(1+z)^2 r(z)a\left(\eta(z),r(z)\right) \,,
\ee
where $\Bigl(t(z),r(z)\Bigr)$ or $\Bigl((\eta(z),r(z)\Bigr)$
is the solution of the radial geodesic equation
as a function of the redshift.

The past-directed radial null geodesic is given by
\bea
\label{geo1}
\frac{dT(r)}{dr}=f(T(r),r) \,;
\quad
f(t,r)=\frac{-R_{,r}(t,r)}{\sqrt{1+2E(r)}} \,.
\eea
where $T(r)$ is the time coordinate along the null radial geodesic as a function of the the coordinate $r$.

From the implicit solution, we can write 
\bea
T(r)=t(U(r),r) \\
\frac{dT(r)}{dr}=\frac{\partial t}{\partial \eta} \frac{dU(r)}{dr}+\frac{\partial t}{\partial r} 
\eea

where $U(r)$ is the $\eta$ coordinate along the null radial geodesic as a function of the the coordinate $r$.

Since it is easier to write down the geodesic equation in the coordinate $(t,r)$ we will start from there 
 \cite{Celerier:1999hp}:
\begin{eqnarray}
{dr\over dz}={\sqrt{1+2E(r(z))}\over {(1+z)\dot {R'}[r(z),t(z)]}} . 
\label{eq:34} \\
\nonumber
{dt\over dz}=-\,{R'[r(z),t(r)]\over {(1+z)\dot {R'}[r(z),t(z)]}} . 
\label{eq:35} \\
\end{eqnarray}

where the $'$ denotes the derivative respect to $r$ and the dot $\dot{}$ the derivative respect to $t$. 
These equations are derived from the definition of redshift and by following the evolution of a short time interval along the null geodesic $T(r)$.

The problem is that there is no exact analytical solution for $R(t,r)$, so the r.h.s. of this equations cannot be evaluated analytically but requires to find a numerical solution for $R$ first \cite{Hellaby:2009vz}
 , and then to integrate numerically the differential equation, which is a quite inconvenient and difficult numerical procedure.

Alternatively a local expansion for $R(t,r)$ around $(t_0,0)$ ,corresponding to the central observer, could be derived and used in eq.(\ref{eq:35}), but being an expansion will loose accuracy as the redshift increases. 

For this reason it is  useful for many numerical and analytical  applications to write the geodesic equations for the coordinates $(\eta,r)$,
\bea
\label{geo3}
\frac{d \eta}{dz}
&=&\frac{\partial_r t(\eta,r)-F(\eta,r)}{(1+z)\partial_{\eta}F(\eta,r)}=p(\eta,r) \,,\\
\label{geo4}
\frac{dr}{dz}
&=&-\frac{a(\eta,r)}{(1+z)\partial_{\eta}F(\eta,r)}=q(\eta,r) \,, \\
F(\eta,r)&=&-\frac{1}{\sqrt{1-k(r)r^2}}\left[\partial_r (a(\eta,r) r)
+\partial_{\eta} (a(\eta,r) r) \partial_r \eta\right]  \, , 
\eea
where $\eta=U(r(z))$ and $F(\eta,r)=f(t(\eta,r),r)$.
It is important to observe that the functions $p,q,F$ have an explicit analytical form which can be obtained from $a(\eta,r)$ and $t(\eta,r)$ as shown below.
 
The derivation of the implicit  solution $a(\eta,r)$ is based on the use of 
the conformal time variable $\eta$, which by construction 
satisfies the relation,
\be
\frac{\partial\eta(t,r)}{\partial t}=a^{-1} \,.
\ee

This means
\bea
t(\eta,r)
&=&t_b(r)+\int^{\eta}_{0}a(\eta^{'},r) d\eta^{'} \, ,
\\
dt&=&a(\eta,r)d\eta+\left(\int^{\eta}_{0}
\frac{\partial a(\eta^{'},r)}{\partial r} d\eta^{'}+t_b^{'}(r)\right) dr \,,
\eea
In order to use the analytical solution we need to find an analytical expression for $F$ and $F_{,\eta}$.

 This can always be done by using
\bea
&& \frac{\partial}{\partial t}=a^{-1}{\frac{\partial}{\partial \eta}} \\
&&\partial_r t(\eta,r)=
\frac{ \rho_0 \, k'(r)}{12 k(r)^{5/2}} 
\left[
3 \sin{ \left( \eta\sqrt{k(r)} \right) }
-\eta \left( 2+\cos{ \left(\eta\sqrt{k(r)}\right) } \sqrt{k(r)} \right)
\right]+t_b'(r)  \, , \\
&&\partial_r \eta=
-a(\eta,r)^{-1}\partial_r t  \, 
\eea 
In this way the coefficients of equations (\ref{geo3}) and (\ref{geo4}) are 
fully analytical, which is a significant improvement over previous 
approaches.
\section{Light-cone metric}
Using this equations we can write the metric on the light-cone in redshift space by using $dr=(dr/dz)dz$ and 

\bea
dt  &=&\frac{dt}{dz} dz=\left(\frac{\partial t}{\partial \eta}\frac{d\eta}{dz}+\frac{\partial t}{dr}\frac{\partial r}{dz}\right)dz \\
p(z)&=&p(\eta(z),r(z)) \\
q(z)&=&q(\eta(z),r(z))  \\
R(z)&=&R(\eta(z),r(z))=r(z)a(\eta(z),r(z))
\eea
to finally get 
\bea
ds_{LC}^2 & = & G(z)dz^2+R(z)^2 d\Omega^2 \\
G(z) & = & q(z)^2 F(\eta(z),r(z))^2-Q(z)^2\\
Q(z) & = & p(z) \partial_{\eta}t(\eta(z),r(z))+q(z) \partial_{r}t(\eta(z),r(z))
\eea

from which we can find directly the radial null geodesic equation for $\eta(z)$ by imposing $G(z)=0$.
This metric only describes space-time in the light-cone, and it is not valid outside, since it is based on the null geodesic congruence given by the radial null geodesics tangent vector field. Since this is the only part of the full LTB space observationally connected to a central observer, it is sufficient to determine models which can explain observational data, which by definition have to be inside the light cone of the central observer.

In principle, given $G(z)$ and $R(z)=D_A(z)$ the light-cone geometry of a LTB is completely defined since this are gauge invariant quantities, but it is not clear what observable to associate to $G(z)$, so the inversion problem still requires to solve the geodesic equations to construct $mn(z)$ and $D_L(z)=(1+z)^2D_A(z)$ .

It should be mentioned that our approach is general while \cite{Mustapha:1997xb} derived an analytical version of the geodesic equation in the light-cone gauge, $t_0-t=-\tilde{r}$, under the assumption $E(r)=0$, just taking into account the inhomogeneity coming from the bang function $t_b(r)$. 
Another related result was obtained by \cite{Nambu:2005zn}, working in the same gauge, 
 where an expression was derived to third order in red-shift for $\tilde{r}(z)$ and $D_L(z)$.

Our results are equivalent after performing the appropriate gauge transformations obtained from the condition

\be
\tilde{M}(\tilde{r})=M(r) \,.
\ee

Such transformations involve the relation between the coefficients of the expansion of $\tilde{M}(\tilde{r}),\tilde{E}(\tilde{r})$ and our $k(r)=-2E(r)/{r^2}$ , which are rather complicated already at third order in red-shift. 
For this reason we will report them separately in a future work in which the general relation between the light-cone gauge and the FLRW gauge is analyzed in detail.

\section{Calculating $mn(z)$}

Expanding the r.h.s. of the geodesics equation we can easily integrate the corresponding polynomial in $q(z),p(z)$, to get $r(z)$ and $\eta(z)$.
It can be easily shown that in order to obtain $D_L(z)$ to the fourth order and $mn(z)$ the fifth  we need to expand $r(z)$ to the fourth order and $\eta(z)$ to the third.

Since we only need $r(z)$ to calculate $mn(z)$ we only give that expansion, but it is understood that $\eta(z)$ has to be computed as well to obtain it.

In order to have an solution which is analytical everywhere we will should use the following expressions for $k(r)$ and $t_b(r)$:

\bea
k(r)&=&k_0+k_2 r^2+k_4 r^4\\
t_b(r)&=&t^b_0+t^b_2 r^2 \,
\eea

which are based on the fact that taking only even powers the functions are analytical everywhere, including the center. We will nevertheless keep $t^b_3$ in the following formulas, in order to provide the general results and show where the differences arise respect to the previous calculations \cite{Nambu:2005zn} to lower order in redshift. 

After re-expressing the results in terms of $H_0$ and $q_0$ we get

\bea
r(z)& = & r_1 z+r_2 z^2+r_3 z^3+r_4 z^4 \nonumber\\
r_1 & = & \frac{1}{H_0} \nonumber\\
r_2 & = & -\frac{q_0+1}{2 H_0} \nonumber\\
r_3 & = & \frac{(2 q_0-1) \left(H_0^4 (1-2 q_0)^2 \left(-\frac{2 q_0 t^b_2}{H_0}+q_0^2+1\right)-5 k_2
   q_0+k_2\right)+6 k_2 \sqrt{2 q_0-1} q_0^2 \arccos\left(\frac{1}{\sqrt{2 q_0}}\right)}{2 H_0^5 (2
   q_0-1)^3} \nonumber \\
r_4 & = &\frac{H_0^4 (1-2 q_0)^2 \left(5 q_0^3-q_0^2+4\right)-2 H_0^3 (1-2 q_0)^2 q_0 (10 q_0+1) t^b_2-3 k_2
   \left(10 q_0^2+5 q_0-2\right)}{8 H_0^5 (2 q_0-1)^2} \nonumber \\
   & &+ \frac{8 H_0^2 q_0 (1-2 q_0)^3 t^b_3+6 k_2 q_0^2 \sqrt{2 q_0-1} (10 q_0+1) \arccos {\left(\frac{1}{\sqrt{2}
   \sqrt{q_0}}\right)}}{8 H_0^5 (2 q_0-1)^3}
\eea

where 
\bea
H_0 & = &\frac{\dot{a}(t_0,0)}{a(t_0,0)} \\
q_0 &= -&\frac{\ddot{a}(t_0,0)\dot{a}(t_0,0)}{\dot{a}(t_0,0)^2}
\eea

The derivative respect to $t$ is denoted with a dot, and is calculated using the analytical solution $a(\eta,r)$ and the derivative respect to $\eta$ is obtained from $\partial_{t}a=\dot{a}=\partial_{\eta}a\, a^{-1}$.

From the definition of $mn(z)$ and the equation for the energy density we can write

\be
4\pi mn(z) dz = \rho d^3 V=\frac{4\pi M'}{\sqrt{1-k(r)r^2}}dr
\ee

from which by using $dr=(dr/dz)dz$ we get

\be
mn(z)=\frac{M'(r(z)}{\sqrt{1-k(r(z))r(z)^2}}\frac{dr(z)}{dz}=\frac{\rho_0 r(z)^2}{2\sqrt{1-k(r(z))r(z)^2}}\frac{dr(z)}{dz}
\ee

where in the last equation we have used the FLRW gauge condition $M(r)=\rho_0 r^3/6$, which allows to calculate $mn(z)$ directly from $r(z)$.

We finally get:

\bea
mn(z) & = & \frac{3 q_0 z^2}{H_0}-\frac{6 q_0 (q_0+1) z^3}{H_0} \nonumber \\
   & & +\frac{3 q_0}{4 H_0^5 (-1 + 2 q_0)^(5/2)}\bigg[\sqrt{2 q_0-1} \left(H_0^4 (1-2 q_0)^2 \left(15 q_0^2+14 q_0+13\right)+10 k_2 (1-5 q_0)\right) \nonumber \\
 & & +20 H_0^3 (1-2 q_0)^2 q_0 \sqrt{2 q_0-1} t^b_2+60 k_2 q_0^2 \arccos\left(\frac{1}{\sqrt{2 q_0}}\right)\bigg]z^4 \nonumber \\
  && + \bigg[\sqrt{2 q_0-1} \left(H_0^4 (1-2 q_0)^2 \left(28 q_0^3+24 q_0^2+21 q_0+19\right)-3 k_2 \left(50 q_0^2+31
   q_0-10\right)\right) \nonumber \\
    & & +\sqrt{2 q_0-1} \left(24 H_0^2 (1-2 q_0)^2 q_0 t^b_3-6 H_0^3 (1-2 q_0)^2 q_0 (14 q_0+5)        t^b_2\right) \nonumber \\
  & &  +18 k_2 (14 q_0+5) q_0^2 \arccos\left(\frac{1}{\sqrt{2 q_0}}\right)\bigg]z^5
\eea

As it can be seen the effects of inhomogeneities show only from the fourth order, and there is no dependency on $k_4$ up to fifth order.

A similar expansion can be obtained for the luminosity distance, and then by combing the two results together we can give a general solution to the inversion problem at low-redshift.

Finally it should be mentioned again that in order to avoid a central singularity we should set $t^b_3=0$.

\section{Relation of $mn(z)$ to observations}
The redshift spherical shell mass $mn(z)$ is the product of the number of sources $n(z)$ times their mass $m(z)$, so it can be related to observations by using the following trivial relation with the total rest mass $M_T(z)$ within a given redshift, i.e. contained in a sphere of comoving radius $r(z)$:

\bea
M_T(z)=\int^z_0 4\pi mn(z')d z' \\
4\pi mn(z)=\frac{M_T(z)}{dz}
\eea

The quantity $M_T(z)$ is obtained from observations by simply adding the mass of all the sources with redshift equal or less than $z$.

Alternatively if we are interested in a more direct relation of $mn(z)$ to observations, without having to use $M_T(z)$, $E_{RSS}(z)$ is the quantity which should be naturally considered, since the uncertainty in the redshift determination would always imply the necessity of same integration in redshift space of $mn(z)$.

The relation between $mn(z)$ and $E_{RSS}(z)$ given in eq.(13,14) is particularly useful for observational purposes, since an appropriate choice of the times scale $\Delta T$ can significantly avoid unwanted astrophysical evolution effects on the sources number counts $n(z)$, which is exactly the reason why $E_{RSS}(z)$ it is defined in that way.

\section{Conclusion}

We have derived a set of differential equations for the radial 
null geodesics in LTB space-time without a cosmological constant and 
applied them to compute the redshift spherical shell mass to fifth order in the red-shift, which could be used to test local inhomogeneities to a higher level of accuracy. 
We have also used the geodesic equations to write the light-con metric in the red-shift space, and show clearly the relation between LTB models and cosmological observables. 

In the future it will be interesting to extended our analysis to other observables such
 as $H(z)$ and $w(z)$ and to use the equations we derived to
 provide a new method to solve the inversion problem of mapping the
 observed luminosity distance $D_L(z)$ and the redshift spherical shell mass $mn(z)$ to LTB models. 
 We can in fact write the differential equation $\partial_z \left(\frac{D_L(z)}{(1+z)^2}\right)=\partial_z \left(r(z)a(\eta(z),r(z)\right) $, where the $D_L(z)$ is the observed luminosity distance and the r.h.s. can be expressed analytically using the geodesic equations we obtained. 
 This equation together with the one we derived for $mn(z)$ should allow to solve for $k(r(z))=k(z)$ and $t_b(r(z))=t_b(z)$, with analytical results at low red-shift and numerical at higher red-shift.
A detailed analysis of such approaches will presented in a future paper.

Anther possible application of our results would be to give an analytical approximation for $E_{RSS}(z)$ \cite{Romano:2007zz}
, the redshift spherical energy, a quantity constructed by integrating $mn(z)$ over varying redshift intervals $\Delta Z(z)$ corresponding to a constant time interval $\Delta t$, which should be the characteristic time scale over which the astrophysical evolution of the astrophysical object counted can be neglected.

The formula we have derived for $mn(z)$ could be used to test local inhomogeneities in red-shift space in a self-consistent way, and it will be the subject of a future paper the corresponding analysis of experimental data from galaxy surveys.

\begin{acknowledgments}
I thank A. Starobinsky and M. Sasaki for useful comments and discussions, and J. Yokoyama for the 
for the hospitality at RESCUE. 
A. E. Romano was supported by JSPS.
This work was also supported in part by JSPS Grant-in-Aid for Scientific 
Research (A) No.~21244033, and by JSPS 
Grant-in-Aid for Creative Scientific Research No.~19GS0219,
and by Monbukagaku-sho Grant-in-Aid for the global COE program,
"The Next Generation of Physics, Spun from Universality and Emergence".

\end{acknowledgments}

\end{document}